%% file: document2.tex
\documentclass[prl,superscriptaddress,twocolumn,showpacs,preprintnumbers,amsmath,amssymb]{revtex4}

\RequirePackage{lineno}  

\usepackage{epsfig}
\usepackage{dcolumn}
\usepackage{bm}

\usepackage{graphicx}

\include{include}

\begin{document}

\title{ \boldmath First Measurement of $ZZ$ Production in $p\bar{p}$
  Collisions at $\sqrt{s}=1.96$ TeV}

\input{authorList/authorList.tex}


\begin{abstract} 

We report the first measurement of the cross section for $\Z$ boson
pair production at a hadron collider. This result is based on a data
sample corresponding to 1.9 ${\rm fb}^{-1}$ of integrated luminosity
from $p\overline{p}$ collisions at $\sqrt{s} = 1.96~{\rm TeV}$
collected with the CDF II detector at the Fermilab Tevatron. In the
$\llll$ channel, we observe three $ZZ$ candidates with an expected
background of $0.096^{+0.092}_{-0.063}$ events. In the $\llnunu$
channel, we use a leading-order calculation of the relative $\ZZ$ and
$\WW$ event probabilities to discriminate between signal and
background. In the combination of $\llll$ and $\llnunu$ channels, we
observe an excess of events with a probability of $5.1\times 10^{-6}$
to be due to the expected background. This corresponds to a
significance of 4.4 standard deviations. The measured cross section is
$\sigma(p\overline{p}\rightarrow \ZZ) = 1.4^{+0.7}_{-0.6}$~(stat.+syst.)~pb, 
consistent with the standard model expectation.
\end{abstract}


\maketitle

\normalsize


Measurements of heavy vector boson pair production ($WW$, $WZ$, $ZZ$)
are of great importance because they test the electroweak sector of
the standard model (SM). Diboson production provides a sensitive probe
of new physics, including anomalous trilinear gauge couplings
\cite{Hagiwara:1986vm}, new particles such as the Higgs boson
\cite{Higgs:1964}, and large extra dimensions \cite{Kober:2007bc}. 
Important tests of electroweak physics have been made recently at the
Fermilab Tevatron with the first observation of $\WW$ production in hadron
collisions \cite{Abazov:2004kc,Acosta:2005mu} and the first
observation of $WZ$ production \cite{Abulencia:2007tu}. The production
of $Z$ pairs has been observed in $e^+e^-$ collisions at LEP
\cite{Alcaraz:2006mx}, but not in hadron collisions. As a window to
new physics, $\ZZ$ production is particularly interesting because of
the absence of $ZZ\gamma$ and $ZZZ$ couplings in the SM, and because
of the very low backgrounds in the four charged-lepton channel. 

The most sensitive previous search for $\ZZ$ production in hadron
collisions was reported by the \Dzero~Collaboration using data
corresponding to 1 ${\rm fb}^{-1}$ of integrated luminosity. That
search used only the four charged-lepton channel and set a limit on
the cross section of $\sigma(ZZ)$ $<$ 4.4 pb at 95\%
C.L. \cite{Abazov:2007xx}. The next-to-leading order (NLO) $\ZZ$ cross
section for $p\overline{p}$ collisions at $\sqrt{s} = 1.96~{\rm TeV}$
is 1.4 $\pm$ 0.1~pb in the zero-width $Z$ approximation
\cite{Campbell:1999ah}.

In this Letter, we report a signal for $\ZZ$ production in hadron
collisions which has a probability of $5.1\times 10^{-6}$ to be due to
the expected background. The production is observed in $p\overline{p}$
collisions at $\sqrt{s} = 1.96~{\rm TeV}$ using a data sample
corresponding to 1.9~${\rm fb}^{-1}$ of integrated luminosity
collected by the CDF II detector. We consider both $\ZZllll$ and
$\ZZllnunu$ decays, where $\ell$ and $\ell^{\prime}$ are electrons or
muons directly from $\Z$ decay or from the leptonic decay of $\tau$'s 
when one or both $\Z$ bosons decay to $\tau$ leptons.


The CDF II detector \cite{Acosta:2004yw} geometry is described using
the azimuthal angle $\phi$ and the pseudo-rapidity $\eta
\equiv-\ln[\tan(\theta/2)]$, where $\theta$ is the polar angle with
respect to the proton beam axis (positive $z$-axis). The
pseudo-rapidity of a particle assumed to originate from the center of
the detector is referred to as $\eta_{\rm{d}}$.

The trajectories of charged particles (tracks) are reconstructed using
silicon micro-strip detectors \cite{SVXII,ISL} and a 96-layer
open-cell drift chamber (COT) \cite{COT} inside a 1.4~T solenoid. A
particle with $\etad \leq 1$ traverses all 96 COT radial
layers. The number of traversed layers decreases to zero as $\etad$
approaches 2. The silicon system provides coverage with 6 (7) layers
with radii between 2.4 cm and 28 cm for
$\etad<1.0~(1.0<\etad<2.0)$. Outside of the solenoid are
electromagnetic (EM) and hadronic (HAD) sampling calorimeters, 
segmented in a projective tower geometry, and constructed of layers of
lead and iron absorber, respectively, and scintillator. The EM section
is the first 19-21 radiation lengths, corresponding to one hadronic
interaction length ($\lambda$), and contains electromagnetic showers.
The HAD section of the calorimeter extends to 4.5-7 $\lambda$ and
contains the majority of a hadronic shower. The calorimeters are
divided into central ($\etad<1.1$) and forward ($1.1<\etad<3.6$)
regions. Outside of the central calorimeters are muon detectors
consisting of scintillators and drift chambers.


The branching fraction of the $\ZZ$ state to four $e$ or $\mu$
leptons, including those from leptonic $\tau$ decays, is only 
0.51\%. When coupled with the small SM cross-section, only a small
number ($\sim$14) of $\ZZllll$ are expected to be produced in
1.9 ${\rm fb}^{-1}$ at the Tevatron. The finite CDF II acceptance
further reduces the expected number of observed $\ZZllll$ events.

To maximize the acceptance, we construct lepton candidates out of all
reconstructed tracks and energy clusters in the EM section of the
calorimeter. This is done with the same lepton identification criteria
used in a previous CDF measurement of $\WZ$ production
\cite{Abulencia:2007tu}. The lepton candidates are divided into seven 
exclusive categories: three for muons, three for electrons, and one
for ``track-only leptons'' which are tracks extrapolating to detector
regions that are inactive for energy measurement because they are
either not covered, or are only partially covered, by calorimeter
components. The three muon categories include one that uses the muon
detectors and two that use different types of minimum ionizing tracks,
central ($\etad<1.1$) and forward ($1.2<\etad<2.0$). The three
electron categories are central ($\etad<1.1$), forward
($1.2<\etad<2.0$) with a matched silicon-based track, and forward
($1.2<\etad<2.8$) without a matched silicon-based track (see
Ref. \cite{Abulencia:2007tu} for more detail).

The transverse energy $\Et$ of an EM cluster or calorimeter tower
is $E\sin\theta$, where $E$ is the associated energy. Similarly,
$p_{T}$ is the component of track momentum transverse to the beam line. 
To suppress jets being misidentified as leptons, all lepton candidates
are required to pass a calorimeter-based isolation requirement such
that the sum of the $E_{T}$ for the calorimeter towers not associated
with the lepton in a cone of $\Delta R = \sqrt{(\Delta\eta)^2 +
(\Delta\phi)^2} < 0.4$ around the lepton direction is less than 10\%
of the $\Et$ for electrons or $\pt$ for muons and track lepton
candidates.


To detect the presence of neutrinos in $\ZZllnunu$ decay, we use
missing transverse energy $\vec{\MET} = - \sum_{i} E_{i} ~
\hat{n}_{T,i}$, where $\hat{n}_{T,i}$ is the transverse component of 
the unit vector pointing from the interaction point to calorimeter
tower $i$. The $\vec{\MET}$ calculation is corrected for muons and
track-only lepton candidates, which do not deposit all of their energy
in the calorimeter. We analogously define the scalar sum $\Sigma\Et =
\sum_{i} E_{T,i}$ applying the same corrections.


The events we consider must pass at least one of four trigger
selection criteria. The central electron trigger requires an EM energy
cluster with $\Et >18$ GeV matched to a track with $\pt >
8~\GeVc$. For the $ZZ\rightarrow ee\nunubar$ channel only, a trigger
for forward electrons requires an EM energy cluster with $\Et >20$ GeV
and an uncorrected, calorimeter-based measurement of $\MET > 15$ GeV.
Two muon triggers are based on stubs from the corresponding muon
detectors matched to a track with $\pt > 18~\GeVc$. Trigger
efficiencies are measured in leptonic $W$ and $Z$ data samples
\cite{Acosta:2004uq}.


The $\ZZllll$ candidates are selected from events with exactly four
charged-lepton candidates using requirements that were optimized with
Monte Carlo simulation without reference to the data. At least one
lepton is required to satisfy the trigger criteria and have $\Et >
20$~GeV ($\pt > 20~\GeVc$) for electrons (muons). We loosen this
requirement to $10$~GeV ($\GeVc$) for the other leptons to increase
the $ZZ$ kinematic acceptance. We require at least two same-flavor,
opposite-sign lepton pairs in the event. Trackless electrons are
considered to have either charge, and track-only leptons either
flavor. One pair must have invariant mass $M_{\ell^{+}\ell^{-}}$ in
the range [76, 106] $\GeV$, while the requirement for the other pair
is extended to [40, 140] $\GeV$ to increase the acceptance for
off-shell $Z$ decays. 

The acceptance for the $\ZZ$ process is determined using {\sc pythia}
\cite{Sjostrand:2006za} Monte Carlo calculations followed by a {\sc
  geant}-based simulation \cite{GEANT} of the CDF II detector. 
An efficiency correction, of up to 10\% per lepton, is applied to the
simulation based on measurements of the lepton reconstruction and
identification efficiencies using observed
$\Z\rightarrow\ell^{+}\ell^{-}$ events.

The dominant backgrounds to the $\ZZllll$ selection are the Drell-Yan
$Z/\gamma^{*}$ process (DY) with two jets misidentified as leptons
($Z$+jets) and DY with an additional photon and a jet, both misidentified 
as leptons ($Z\gamma$+jets). The $Z$+jets and $Z\gamma$+jets
background contributions are estimated from data by extrapolating from
a sample of events that contain three identified leptons and a jet
$j_l$ containing a track or EM energy cluster similar to those
required in the lepton identification. The contribution of each event
to the total yield is scaled by the probability that the $j_l$ is
identified as a lepton. This probability $p(j_l)$ is determined from
multijet events collected with jet-based triggers and is a function of
the $j_l$ $p_T$ and type of lepton. A correction to $p(j_l)$ is
applied for the small real lepton contribution using Monte Carlo
simulation of single $\W$ and $\Z$ boson processes.  In this
background sample, one of the three identified leptons is likely to be
either a jet or a photon misidentified as a lepton. An event with
two leptons and two $j_l$ jets enters the three lepton plus $j_l$
sample if either of the $j_l$ jets is misidentified as a lepton, but
will enter the four identified lepton sample only if both $j_l$ 
jets are misidentified as leptons. Therefore, the contribution from
this category of events is double counted. A correction for this is
made by subtracting the yield of two leptons plus two $j_l$ jets
scaled by $p(j_{l,1})\times p(j_{l,2})$. As the three lepton plus
$j_l$ sample has significant contributions from the $\ZZllll$ signal
itself when one of the leptons is not fully identified but is counted
as a $j_l$, an anti-isolation requirement of $>$20\% (see previous
definition of calorimeter-based isolation) is applied to the $j_l$
selection.

As a cross-check, we estimate the $Z\gamma$+jet background contribution
in an alternative way using the yield of three lepton plus $j_l$
events in simulated $Z\gamma$+jet data scaled by $p(j_l)$. We correct
this with the ratio of data to simulation for an analogous calculation 
of two identified leptons and one $j_l$ scaled by $p(j_l)$ in $Z$+jets
events. This estimate of the $Z\gamma$+jet background is in good
agreement with the nominal estimate based solely on the data.


We separate the $\ZZllll$ candidates into two exclusive categories
based on whether or not they contain at least one forward electron
without a track. This is done because the background from
$Z\gamma$+jets is much larger in candidates with a forward trackless
electron. The expected signal and background yields assuming
$\sigma(ZZ) = 1.4~\pm$ 0.1~pb and the observed yields are shown in
Table~\ref{tbl:ZZllll_results}.


The statistical significance of the $\ZZllll$ yield is determined
using a maximum likelihood fit with two bins, one for each of the
$\ZZllll$ categories. We define ${\Delta\ln\cal L}$ as the logarithm
of the likelihood ratio between this fit and the no signal
hypothesis. In $10^{7}$ background-only Monte Carlo experiments, only
109 have larger ${\Delta\ln\cal L}$ than that observed in data. This
corresponds to a background-only probability (p-value) of  $1.1 \times
10^{-5}$ and a signal significance equivalent to 4.2 standard deviations.

\begin{table}
\caption{Expected and observed number of $\ZZllll$ candidate events.
  The first uncertainty is statistical and the second one is systematic. 
  }
\label{tbl:ZZllll_results}
\begin{ruledtabular}
\begin{tabular}{lcc}
               & Candidates without a                             &     Candidates with a              \\
Category       & trackless electron                             &     trackless electron           \\  
               
\hline         
   $ZZ$             &   1.990 $\pm$ 0.013 $\pm$ 0.210           &     0.278 $\pm$ 0.005 $\pm$ 0.029      \\
   $Z$+jets       \parbox{0.0cm}{\vspace{0.55cm}}   &   ${0.014}^{+0.010}_{-0.007} \pm 0.003$   &  ${0.082}^{+0.089}_{-0.060} \pm 0.016$ \\
\hline
   Total          \parbox{0.0cm}{\vspace{0.65cm}}  &   $2.004^{+0.016}_{-0.015} \pm 0.210 $    &  $0.360^{+0.089}_{-0.060} \pm 0.033$   \\
\hline
{\bf Observed}      &             {\bf  2 }                           &      {\bf  1 }     \\
\end{tabular}
\end{ruledtabular}
\end{table}


The $\ZZllnunu$ candidates are selected from events with exactly two
lepton candidates excluding events with forward electrons without
a track which are contaminated by large $W\gamma$ backgrounds. At
least one lepton is required to satisfy the trigger and have $\Et >
20$~GeV ($\pt > 20~\GeVc$) for electrons (muons). This requirement is
loosened to $10$~GeV ($\GeVc$) for the other lepton. We apply a
track-based isolation selection in which the sum of the $p_T$ of the
tracks not associated with the lepton within a cone of $\Delta R <
0.4$ around the lepton is required to be less than 10\% of the
momentum of the track associated with the lepton.


Aside from $ZZ$ production, other SM processes that can lead to two
high-$\pt$ leptons include events from DY, a $\W$ decay with photon
($W\gamma$) or jet ($W$+jets) misidentified as a lepton; and $\ttbar$,
$\WW$, and $\WZ$ production. The $\ttbar$ contribution is suppressed
by requiring fewer than two reconstructed jets with $\Et > 15$ GeV and
$|\eta_d| < 2.5$ in the event. The DY background is suppressed by
requiring sufficiently large $\MET$ in the event to remove
contributions from mismeasured leptons and/or jets. This is achieved
by requiring $\METrel > 25~\mathrm{GeV}$, where  
\begin{equation}
  \METrel \equiv \left\{ 
  \begin{array}{ll} 
    \MET                       & \mbox{if } \MetDeltaPhi > \frac{\pi}{2} \\
    \MET\sin{\MetDeltaPhi}   & \mbox{if } \MetDeltaPhi < \frac{\pi}{2} \\
  \end{array} \right\}
\end{equation}
and $\MetDeltaPhi$ is the angle between the $\vec{\MET}$ direction
and the nearest lepton or jet. To suppress events with $\MET$ from
mismeasured unclustered energy, we require significant $\MET$ such
that $\MET/\sqrt(\Sigma\Et) > 2.5$ GeV$^{1/2}$. We require the lepton
pair to be consistent with the same-flavor, opposite-sign property of
$Z$ decay and have dilepton invariant mass $M_{\ell^{+}\ell^{-}} >
16~\GeV$ to suppress QCD backgrounds.

The acceptances for the $\WW$, $\WZ$, $\ZZ$, $W\gamma$, and $\ttbar$
processes are determined using the same detector simulation as
described for the $\ZZllll$ channel. Events are simulated with the
{\sc MC@NLO} program for $\WW$ \cite{Frixione:2002ik}, {\sc pythia}
for $\WZ$, $\ZZ$, and $\ttbar$ \cite{Sjostrand:2006za}, and the
generator described in \cite{Baur:1992cd} for $W\gamma$. An additional
correction is applied to the $W\gamma$ background estimate based on a
measurement of the photon conversion veto efficiency in data. The
background from $W$+jets is estimated from the yield of one identified
lepton plus one $j_l$ scaled by $p(j_l)$. As the sample size is
sufficiently large, a loose calorimeter-based isolation cut ($<$ 30\%)
is applied to the $j_l$ samples to reduce the magnitude of the
extrapolation from the $j_l$ to the fully-identified lepton.

\begin{figure}
\includegraphics[width=0.33\textwidth]{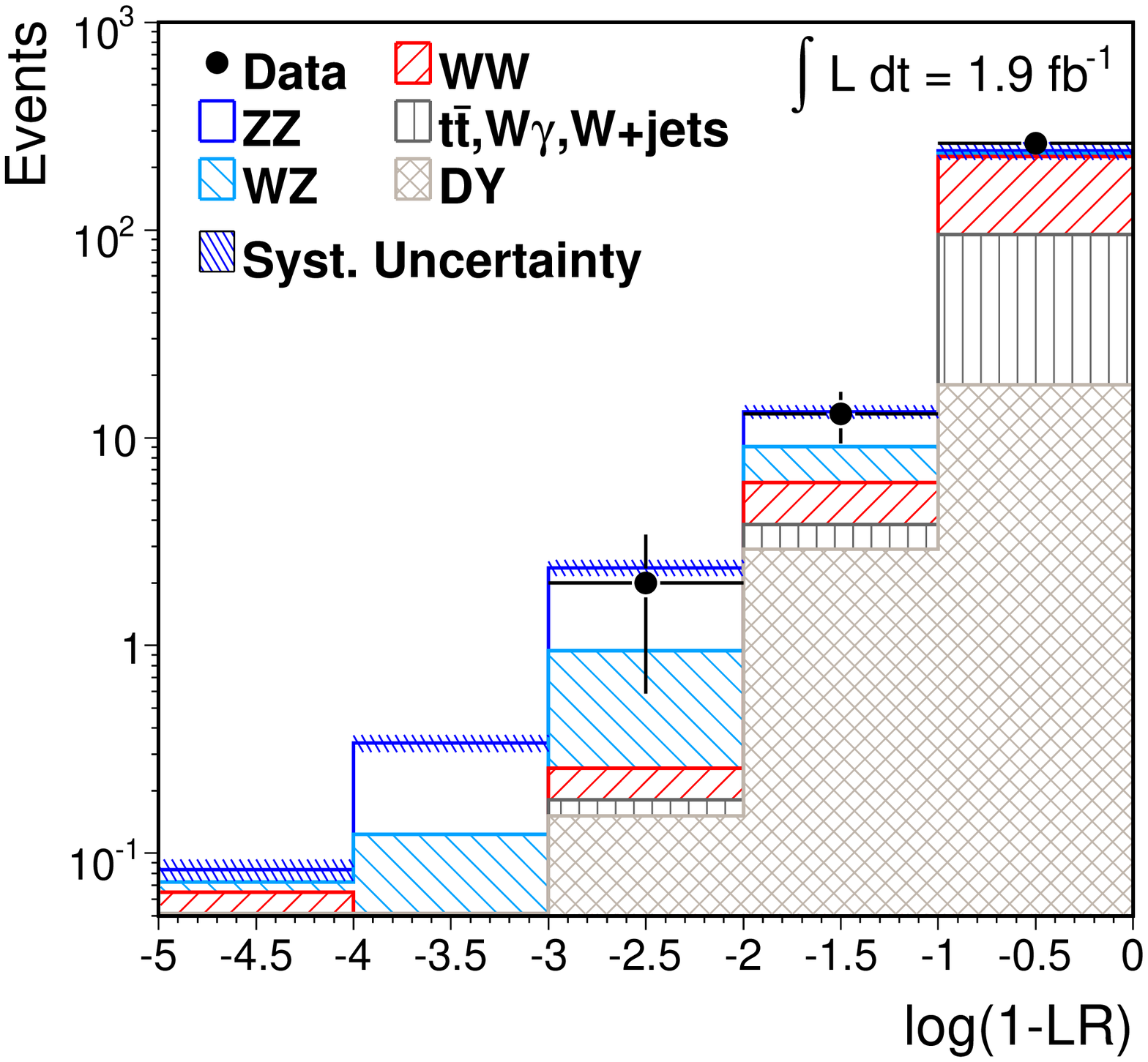}
\caption{Distribution of the discriminating variable $\log_{10}
  (1-LR)$ for the $\ZZllnunu$ search.}
\label{fig:logLR}
\end{figure}

We observe 276 events in the selected region which is expected to
contain $256\pm21$ events of which only $14\pm2$ are from the
$\ZZllnunu$ process in the SM. Approximately half of the yield is due
to the $WW$ process. However, $\ZZllnunu$ and $\WW$ have different
kinematic properties which are exploited to statistically separate the
contribution of these two processes to the dataset. We calculate an
event-by-event probability density for the observed lepton momenta and
$\MET$ using leading order calculations of the differential decay rate
for the processes \cite{Campbell:1999ah}. The event probability
density is 
\begin{equation}
\label{eqn:master_prob}
P(x_{obs}) =
  {1 \over \langle\sigma\rangle }
  \int \frac {d \sigma_{LO} (y) }{ dy }
  \epsilon (y) G(x_{obs},y) dy,
\end{equation}
where the elements of $y$ ($x_{obs}$) are the true (observed) values
of the lepton momenta and $\MET$, $d \sigma_{LO} \over dy$ is the
parton level cross section differential in those observables,
$\epsilon(y)$ is the detector acceptance and efficiency function, and
$G(x_{obs},y)$ is the transfer function representing the detector
resolution. The constant $\langle\sigma\rangle$ normalizes the total
event probability to unity. The missing information due to the fact
that we have two neutrinos in the final state is integrated over in
this calculation. We then form a likelihood ratio discriminant $LR$
which is the signal probability divided by the sum of signal and
background probabilities $LR=P_{ZZ}/(P_{ZZ}+P_{WW})$. The distribution
of $\log_{10} (1-LR)$ for the data compared to the summed signal and
background expectation is shown in Figure~\ref{fig:logLR}.

For both $\llll$ and $\llnunu$ channels, the systematic uncertainties
associated with the  Monte Carlo simulation affect the $WW$,  $WZ$,
$ZZ$, $W\gamma$, and $\ttbar$ expectations similarly. The
uncertainties from the lepton selection and trigger efficiency
measurements are propagated through the analysis, giving uncertainties
from $1.0\%$ to $1.4\%$ and $2.1\%$ to $6.1\%$ for the respective
efficiencies of the different signal and background processes. The
detector acceptance variation due to PDF uncertainties is assessed to
be in the range $1.9\%$ - $4.1\%$ using the 20 pairs of PDF sets
described in \cite{Kretzer:2003it}.  We assign a 6\% luminosity
uncertainty to signal and background estimates obtained from
simulation \cite{Acosta:2002hx}. The cross-section uncertainties are
10\% for $WW$ \cite{Campbell:1999ah}, $WZ$ \cite{Campbell:1999ah}, and
$W\gamma$ \cite{Baur:1997kz}, and 15\% for $\ttbar$
\cite{Kidonakis:2003qe,Cacciari:2003fi}.

The systematic uncertainty on the $W$+jets background to $\ZZllnunu$
candidates and the $Z$+jets and $Z\gamma$+jets background to the
$\ZZllll$ is estimated to be 20\% from differences in the observed
probability that a jet is identified as a lepton for jets collected
using different jet $\Et$ trigger thresholds. These variations
correspond to changing the parton composition of the jets and the
relative amount of contamination from real leptons. 

For the $\ZZllnunu$, the systematic uncertainty on the DY background
yield due to the $\MET$ resolution modeling is estimated to be 20\%
from comparisons of the data and Monte Carlo simulation in a sample of
dilepton events. For the $W\gamma$ background contribution, there is
an additional uncertainty of 20\% from the detector material
description and conversion veto efficiency. 


We define four independent control samples based on the $\ZZllnunu$
selection where one of the cuts is removed or inverted to test our
modeling of background-dominated data. Removing the $\MET$ requirement
produces a DY-dominated sample which tests luminosity accounting,
lepton reconstruction efficiency, and non-$\MET$ related trigger
efficiencies that apply to both the $\llll$ and $\llnunu$ final
states; we observe (expect) 160980 ($160000\pm18000$) events, where
the uncertainty combines statistical and systematic
contributions. Inverting the charge sign requirement to select 
same-sign dilepton events tests our modeling of photons and jets
misidentified as leptons; we observe (expect) 161 ($138\pm19$) events.
Inverting the $\MET/\sqrt(\Sigma\Et)$ requirement but selecting
$\METrel > 25$ GeV events tests our modeling of the effect of
unclustered energy on the $\MET$; we observe (expect) 55 ($59\pm9$)
events. Finally, inverting the $\METrel>25$ GeV requirement but
selecting $\MET > 25$ GeV events tests our modeling of mismeasured
leptons or jets at high $\MET$; we observe (expect) 151 ($178\pm30$)
events. The observed yields are in good agreement with the
expectations in each selection set. 


We combine the $\ZZllnunu$ with the $\ZZllll$ results by extending the
likelihood fit previously described to include the $\log_{10} (1-LR)$
distribution. The p-value for the $\ZZllnunu$ alone is 0.12 and the
combined p-value is $5.1\times10^{-6}$ corresponding to a significance
equivalent to 4.4 standard deviations. We determine a $ZZ$
cross-section for this signal by fitting the data for the fraction of
the expected SM yield in the full acceptance and scaling the
zero-width $Z$ boson approximation cross-section by that fraction. The
measured cross section is  
$\sigma(p\overline{p}\rightarrow \ZZ) =  1.4^{+0.7}_{-0.6}~\mathrm{pb},$
consistent with the SM expectation. This is the first measurement of
the $\ZZ$ cross section in hadron collisions.


We thank the Fermilab staff and the technical staffs of the
participating institutions for their vital contributions. This work
was supported by the U.S. Department of Energy and National Science
Foundation; the Italian Istituto Nazionale di Fisica Nucleare; the
Ministry of Education, Culture, Sports, Science and Technology of
Japan; the Natural Sciences and Engineering Research Council of
Canada; the National Science Council of the Republic of China; the
Swiss National Science Foundation; the A.P. Sloan Foundation; the
Bundesministerium f\"ur Bildung und Forschung, Germany; the Korean
Science and Engineering Foundation and the Korean Research Foundation;
the Science and Technology Facilities Council and the Royal Society,
UK; the Institut National de Physique Nucleaire et Physique des
Particules/CNRS; the Russian Foundation for Basic Research; the
Comisi\'on Interministerial de Ciencia y Tecnolog\'{\i}a, Spain; the
European Community's Human Potential Programme; the Slovak R\&D Agency;
and the Academy of Finland. 

\bibliography{document2}

\end{document}

%% file: include.tex
\def\simge{\mathrel{%
   \rlap{\raise 0.511ex \hbox{$>$}}{\lower 0.511ex \hbox{$\sim$}}}}
\def\simle{\mathrel{
   \rlap{\raise 0.511ex \hbox{$<$}}{\lower 0.511ex \hbox{$\sim$}}}}

\newcommand{\pt}{p_T}
\newcommand{\Et}{E_T}

\newcommand {\GeVc} {{\rm GeV}/c}

\newcommand {\GeV} {{\rm GeV}/c^2}

\newcommand{\MET}{\mbox{$E\kern-0.60em\raise0.10ex\hbox{/}_{T}$}}
\newcommand{\METrel}{\mbox{$E\kern-0.60em\raise0.10ex\hbox{/}_{T,rel}$}}

\newcommand{\W}{W}
\newcommand{\Z}{Z}
\newcommand{\WW}{WW}
\newcommand{\WZ}{WZ}
\newcommand{\ZZ}{ZZ}

\newcommand{\ttbar}{t \overline{t}}

\newcommand{\Dzero} {\rm{D}\O}

\newcommand{\etad}{|\eta_{\rm{d}}|}

\newcommand{\MetDeltaPhi}{\Delta\phi_{E\kern-0.50em\raise0.08ex\hbox{\scriptsize /}_{T},(\ell{\rm,jet)}}}

\newcommand{\nunubar}{\nu\nu}

\newcommand{\llll}{\ell\ell\ell^{\prime}\ell^{\prime}}
\newcommand{\llnunu}{\ell\ell\nunubar}
\newcommand{\ZZllll}{\ZZ \rightarrow \llll}
\newcommand{\ZZllnunu}{\ZZ \rightarrow \llnunu}

%% file: authorList/authorList.tex
\affiliation{Institute of Physics, Academia Sinica, Taipei, Taiwan 11529, Republic of China} 
\affiliation{Argonne National Laboratory, Argonne, Illinois 60439} 
\affiliation{Institut de Fisica d'Altes Energies, Universitat Autonoma de Barcelona, E-08193, Bellaterra (Barcelona), Spain} 
\affiliation{Baylor University, Waco, Texas  76798} 
\affiliation{Istituto Nazionale di Fisica Nucleare, University of Bologna, I-40127 Bologna, Italy} 
\affiliation{Brandeis University, Waltham, Massachusetts 02254} 
\affiliation{University of California, Davis, Davis, California  95616} 
\affiliation{University of California, Los Angeles, Los Angeles, California  90024} 
\affiliation{University of California, San Diego, La Jolla, California  92093} 
\affiliation{University of California, Santa Barbara, Santa Barbara, California 93106} 
\affiliation{Instituto de Fisica de Cantabria, CSIC-University of Cantabria, 39005 Santander, Spain} 
\affiliation{Carnegie Mellon University, Pittsburgh, PA  15213} 
\affiliation{Enrico Fermi Institute, University of Chicago, Chicago, Illinois 60637} 
\affiliation{Comenius University, 842 48 Bratislava, Slovakia; Institute of Experimental Physics, 040 01 Kosice, Slovakia} 
\affiliation{Joint Institute for Nuclear Research, RU-141980 Dubna, Russia} 
\affiliation{Duke University, Durham, North Carolina  27708} 
\affiliation{Fermi National Accelerator Laboratory, Batavia, Illinois 60510} 
\affiliation{University of Florida, Gainesville, Florida  32611} 
\affiliation{Laboratori Nazionali di Frascati, Istituto Nazionale di Fisica Nucleare, I-00044 Frascati, Italy} 
\affiliation{University of Geneva, CH-1211 Geneva 4, Switzerland} 
\affiliation{Glasgow University, Glasgow G12 8QQ, United Kingdom} 
\affiliation{Harvard University, Cambridge, Massachusetts 02138} 
\affiliation{Division of High Energy Physics, Department of Physics, University of Helsinki and Helsinki Institute of Physics, FIN-00014, Helsinki, Finland} 
\affiliation{University of Illinois, Urbana, Illinois 61801} 
\affiliation{The Johns Hopkins University, Baltimore, Maryland 21218} 
\affiliation{Institut f\"{u}r Experimentelle Kernphysik, Universit\"{a}t Karlsruhe, 76128 Karlsruhe, Germany} 
\affiliation{Center for High Energy Physics: Kyungpook National University, Daegu 702-701, Korea; Seoul National University, Seoul 151-742, Korea; Sungkyunkwan University, Suwon 440-746, Korea; Korea Institute of Science and Technology Information, Daejeon, 305-806, Korea; Chonnam National University, Gwangju, 500-757, Korea} 
\affiliation{Ernest Orlando Lawrence Berkeley National Laboratory, Berkeley, California 94720} 
\affiliation{University of Liverpool, Liverpool L69 7ZE, United Kingdom} 
\affiliation{University College London, London WC1E 6BT, United Kingdom} 
\affiliation{Centro de Investigaciones Energeticas Medioambientales y Tecnologicas, E-28040 Madrid, Spain} 
\affiliation{Massachusetts Institute of Technology, Cambridge, Massachusetts  02139} 
\affiliation{Institute of Particle Physics: McGill University, Montr\'{e}al, Canada H3A~2T8; and University of Toronto, Toronto, Canada M5S~1A7} 
\affiliation{University of Michigan, Ann Arbor, Michigan 48109} 
\affiliation{Michigan State University, East Lansing, Michigan  48824} 
\affiliation{University of New Mexico, Albuquerque, New Mexico 87131} 
\affiliation{Northwestern University, Evanston, Illinois  60208} 
\affiliation{The Ohio State University, Columbus, Ohio  43210} 
\affiliation{Okayama University, Okayama 700-8530, Japan} 
\affiliation{Osaka City University, Osaka 588, Japan} 
\affiliation{University of Oxford, Oxford OX1 3RH, United Kingdom} 
\affiliation{University of Padova, Istituto Nazionale di Fisica Nucleare, Sezione di Padova-Trento, I-35131 Padova, Italy} 
\affiliation{LPNHE, Universite Pierre et Marie Curie/IN2P3-CNRS, UMR7585, Paris, F-75252 France} 
\affiliation{University of Pennsylvania, Philadelphia, Pennsylvania 19104} 
\affiliation{Istituto Nazionale di Fisica Nucleare Pisa, Universities of Pisa, Siena and Scuola Normale Superiore, I-56127 Pisa, Italy} 
\affiliation{University of Pittsburgh, Pittsburgh, Pennsylvania 15260} 
\affiliation{Purdue University, West Lafayette, Indiana 47907} 
\affiliation{University of Rochester, Rochester, New York 14627} 
\affiliation{The Rockefeller University, New York, New York 10021} 
\affiliation{Istituto Nazionale di Fisica Nucleare, Sezione di Roma 1, University of Rome ``La Sapienza," I-00185 Roma, Italy} 
\affiliation{Rutgers University, Piscataway, New Jersey 08855} 
\affiliation{Texas A\&M University, College Station, Texas 77843} 
\affiliation{Istituto Nazionale di Fisica Nucleare, University of Trieste/\ Udine, Italy} 
\affiliation{University of Tsukuba, Tsukuba, Ibaraki 305, Japan} 
\affiliation{Tufts University, Medford, Massachusetts 02155} 
\affiliation{Waseda University, Tokyo 169, Japan} 
\affiliation{Wayne State University, Detroit, Michigan  48201} 
\affiliation{University of Wisconsin, Madison, Wisconsin 53706} 
\affiliation{Yale University, New Haven, Connecticut 06520} 
\author{T.~Aaltonen}
\affiliation{Division of High Energy Physics, Department of Physics, University of Helsinki and Helsinki Institute of Physics, FIN-00014, Helsinki, Finland}
\author{J.~Adelman}
\affiliation{Enrico Fermi Institute, University of Chicago, Chicago, Illinois 60637}
\author{T.~Akimoto}
\affiliation{University of Tsukuba, Tsukuba, Ibaraki 305, Japan}
\author{M.G.~Albrow}
\affiliation{Fermi National Accelerator Laboratory, Batavia, Illinois 60510}
\author{B.~\'{A}lvarez~Gonz\'{a}lez}
\affiliation{Instituto de Fisica de Cantabria, CSIC-University of Cantabria, 39005 Santander, Spain}
\author{S.~Amerio}
\affiliation{University of Padova, Istituto Nazionale di Fisica Nucleare, Sezione di Padova-Trento, I-35131 Padova, Italy}
\author{D.~Amidei}
\affiliation{University of Michigan, Ann Arbor, Michigan 48109}
\author{A.~Anastassov}
\affiliation{Rutgers University, Piscataway, New Jersey 08855}
\author{A.~Annovi}
\affiliation{Laboratori Nazionali di Frascati, Istituto Nazionale di Fisica Nucleare, I-00044 Frascati, Italy}
\author{J.~Antos}
\affiliation{Comenius University, 842 48 Bratislava, Slovakia; Institute of Experimental Physics, 040 01 Kosice, Slovakia}
% ...
\collaboration{CDF Collaboration\footnote{With visitors from $^a$University of Athens, 15784 Athens, Greece, 
$^b$Chinese Academy of Sciences, Beijing 100864, China, 
$^c$University of Bristol, Bristol BS8 1TL, United Kingdom, 
$^d$University Libre de Bruxelles, B-1050 Brussels, Belgium, 
$^e$University of California Irvine, Irvine, CA  92697, 
$^f$University of California Santa Cruz, Santa Cruz, CA  95064, 
$^g$Cornell University, Ithaca, NY  14853, 
$^h$University of Cyprus, Nicosia CY-1678, Cyprus, 
$^i$University College Dublin, Dublin 4, Ireland, 
$^j$University of Edinburgh, Edinburgh EH9 3JZ, United Kingdom, 
$^k$University of Heidelberg, D-69120 Heidelberg, Germany, 
$^l$Universidad Iberoamericana, Mexico D.F., Mexico, 
$^m$University of Manchester, Manchester M13 9PL, England, 
$^n$Nagasaki Institute of Applied Science, Nagasaki, Japan, 
$^o$University de Oviedo, E-33007 Oviedo, Spain, 
$^p$Queen Mary, University of London, London, E1 4NS, England, 
$^q$Texas Tech University, Lubbock, TX  79409, 
$^r$IFIC(CSIC-Universitat de Valencia), 46071 Valencia, Spain, 
}}
\noaffiliation